\documentstyle[aps,prl,manuscript]{revtex}
\begin{document}
\draft
%%%%%%%%%%%%%%%%%%%%%%%%%%%%%%%%%%%%%%%%%%%%%%%%%%%%%%%%%%%%%%%%%%%
\newcommand{\vektor}[1]{\mbox{\boldmath $#1$}}
%%%%%%%%%%%%%%%%%%%%%%%%%%%%%%%%%%%%%%%%%%%%%%%%%%%%%%%%%%%%%%%%%%%
\title{DMRG and the Two Dimensional t-J Model}
\author{I. P. McCulloch${}^{\rm 1}$, A. R. Bishop${}^{\rm 2}$
and M. Gulacsi${}^{\rm 1}$}
\address{
${}^{\rm 1}$
Department of Theoretical Physics, Institute of Advanced Studies \\
The Australian National University, Canberra, ACT 0200, Australia \\
${}^{\rm 2}$
Theoretical Division, Los Alamos National Laboratory \\
Los Alamos, NM 87545, U.S.A.}
%% \date{\today}
\maketitle
\begin{abstract}

We describe in detail the application of the
recent non-Abelian Density Matrix Renormalization Group (DMRG) algorithm
to the two dimensional t-J model.
This extension of the DMRG algorithm allows us to keep the equivalent of
twice as many basis states as the conventional DMRG algorithm for the
same amount of computational effort, which permits a deeper understanding
of the nature of the ground state.

\end{abstract}
\pacs{ ~ }

Since the discovery of superconductivity in the rare-earth copper
oxides there has been a growing interest in strongly correlated electronic
systems. The t-J model proposed by Anderson (1987), and Zhang and
Rice (1988) is an example of this interest. The theoretical predictions and
implications of the model are possibly relevant and useful for a deeper
understanding, particularly, the high temperature superconductors,
and in generally the motion of "holes" in an antiferromagnet.

It is conceptually a simple model and has been already widely studied
in relation to the high temperature superconductors. However,
it belongs to the class of systems which do not obey the condition
of Perron-Frobenuis (Yosida 1980). This condition states that if the
off-diagonal elements of a matrix are all non-positive and if the
matrix is not in a block diagonal form then the ground state
eigenvalue is non-degenerate. In the case of the t-J Hamiltonian
the off-diagonal elements are not all non-positive. Thus the above
theorem can not be applied, which implies that the
phenomenon of ground state level crossing is present
(Itoyama, McCoy and Perk 1990). As a direct consequence of this,
the thermodynamic system will be unstable against phase separation.
Indeed, as is known (Blatt and Weisskopf 1979), in three-dimension,
the system will collapse to arbitrarily high density. In two-dimensions
we have the argument of Emery, Kivelson and Lin (1990) that the model
is unstable against phase separation into hole-rich and no-hole phase.

It is known that interchanging particles of different spin
leads to a strong coupling between the kinetic and spin degrees of
freedom(Barnes 1989). Therefore, in different dimensions the model
will represent different cases. Phase separation occurs also in one
dimension, but for different
values of the band filling and interaction strength than in the higher
dimension case. In two dimensions, it was argued that phase separation
corresponds to stripe formation (Hellberg and Manousakis 1999). Stripe
formation is one of the most controversial issues in the study of high
temperature superconductors, where there is a phase separation of the
holes which is limited to short range by, e.g., Coulomb forces, and
orders in striped structures. Many experiments have
found evidence for stripes \footnote{For a recent review, see
the proceedings of the {\it Conference on Spectroscopies in Novel
Superconductors} (Proceedings, 1998).} and they arise
in a number of different theories involving strong correlations
(Zaanen and Gunnarsson 1989; Kivelson, Emery and Lin 1990; Zaanen,
Horbach and van Saarloos 1996; White and Scalapino 1998).

The first direct proof of the existence of a stripe ground state
emerged from t-J model calculations (White and Scalapino 1998).
The reason for this is that the t-J model inherits all the exchange
hole correlations resulting from the antiparallel spin correlations
in the $U \rightarrow \infty$ limit of the Fermi sea. Hence, if stripes
do exist, they will be more robust in the t-J model. Stripes have been
studied by a number of numerical techniques in the t-J model,
unfortunately resulting in conflicting conclusions. A major
question is if stripes are part of the known phase separated regime
of the t-J model (Hellberg and Manousakis 1999) or they represent
a different ground state. This is the focus of our study.

Using the recent developed non-Abelian DMRG (McCulloch and Gulacsi 2000)
we study the ground state of the two dimensional t-J model in the
vicinity of $J/t = 0.35$ and for different size square lattices up to 24x6.
We clearly find, for our choice of boundary conditions, a striped
ground state. We have also noticed that in the phase separated regimes,
the holes are attracted to the open boundary, while in the stripe
regime, the holes are repelled by the boundary. This is evidence
that that stripes are different from a usual phase transition
phenomena, but also indicates that there are large finite size
effects arising from the existence of the open boundary.

The non-Abelian DMRG algorithm
is numerically similar to the ``Interaction Round a Face'' (IRF)
DMRG (Sierra and Nishino 1997). However, the decomposition of
the Hamiltonian into the blocks used by the DMRG algorithm
is significantly different between the two algorithms. In
IRF-DMRG, the vertex Hamiltonian is first transformed
into an IRF model, and then a variant of DMRG is applied
to the IRF Hamiltonian. In the non-Abelian case, the
standard DMRG algorithm is generalized so that the block
operators transform as irreducible representations of an arbitrary
compact group (in this case $SU(2)$). This is a true generalization
of the original DMRG algorithm (White 1992, 1993), in the sense
that if the global symmetry group is chosen to be $U(1)$, as in
the usual $S^z$ basis used in DMRG, the original DMRG algorithm
is recovered exactly. In this paper, we fill in the details
required to apply the non-Abelian algorithm
to the 2 dimensional $t-J$ model, in the basis given by particle number,
$N$, and total spin, $j$, giving the global symmetry group
$U(1) \otimes SU(2)$. We then compare the results of this choice
of basis with previous calculations using the usual $S^z$ basis,
and comment on the conclusions that can be inferred
from the DMRG calculations.

The $t-J$ Hamiltonian is
\begin{equation}
H = - t \sum_{\langle i,j \rangle, \sigma}
(c^\dagger_{i\sigma} c_{j\sigma} + \mbox{H.c.})
+ J (\sum_{\langle i,j \rangle} \vektor{S}_i \cdot \vektor{S}_j
- \frac{1}{4}  n_i n_j)
\end{equation}
defined on the subspace of no double occupancy, and $\langle i,j \rangle$
means summation over nearest neighbour pairs only.  The single site
operators act on the usual three dimensional t-J basis of an empty
site (hole), a single up spin and a single down spin. The difference
for non-Abelian DMRG is that the spin up and spin down states form an
$SU(2)$ multiplet of dimension 2, thus reducing the single site basis
to two states; a hole (transforming as the $[0,0]$ representation of
$U(1) \otimes SU(2)$) and a single spin (transforming as the $[1,1/2]$
representation of $U(1) \otimes SU(2)$). All the single site operators
are represented as $2 \times 2$ matrices acting on these states. The
block basis states used in each step of the DMRG iteration are
labelled $\vert \vert n j (\alpha) \rangle$, which denotes the
$\alpha$'th state of $n$ particles and total spin $j$. The matrix
elements of the single site operators are simply the {\em reduced
matrix elements} given by the Wigner-Eckart theorem
(Biedenharn and Louck 1981),
written here in our $U(1) \otimes SU(2)$ basis,
\begin{equation}
\langle n^\prime j^\prime m^\prime (\alpha^\prime)
\vert T^J_M \vert n j m (\alpha) \rangle \: = \:
C^{j J j^\prime}_{m M m^\prime}
\langle n^\prime j^\prime (\alpha^\prime) \vert \vert
\vektor{T}^J \vert \vert n j (\alpha) \rangle \; ,
\label{eq:WignerEckart}
\end{equation}
for the $M$'th component of an operator $\vektor{T}^J$ transforming
as the $[J]$ representation of $SU(2)$.
$C^{j^{} J j^{\prime}}_{m^{} M m^{\prime}}$ denotes the
Clebsch-Gordan coefficient. We have suppressed the trivial $U(1)$
label on the $\vektor{T}^J$ operator. This theorem specifies
the relationship between the reduced matrix elements acting on
the 2-state $U(1) \otimes SU(2)$ basis, and the 3-state
$U(1) \otimes U(1)$ basis of particle number and $z-$component
of spin \footnote{In a real calculation using the $U(1) \otimes U(1)$
basis, the $j$ index on the left hand side of Eq.\ (\ref{eq:WignerEckart})
cannot be used as a good quantum number since the usual DMRG density matrix
in this basis does not preserve any non-Abelian global symmetries.}.

In matrix form, choosing basis vectors $(1,0)$ to be a hole,
and $(0,1)$ to be a spin, The operators relevant to the t-J model are
\begin{eqnarray}
c^{[-1,1/2]} & = & \left( \begin{array}{cc} 0 & \sqrt{2} \\
0 & 0 \end{array} \right) \nonumber \\
c^{\dagger[1,1/2]} & = & \left( \begin{array}{cc} 0 & 0 \\
1 & 0 \end{array} \right) \nonumber \\
s^{[0,1]} & = & \left( \begin{array}{cc} 0 & 0 \\
0 & \sqrt{3/4} \end{array} \right) \nonumber \\
n^{[0,0]} & = & \left( \begin{array}{cc} 0 & 0 \\
0 & 1 \end{array} \right) \nonumber \\
p^{[0,0]} & = & \left( \begin{array}{cc} 1 & 0 \\
0 & -1 \end{array} \right) \nonumber \; ,
\end{eqnarray}
where we use square brackets to denote which representation of
$U(1) \otimes SU(2)$ the operators transform as. $p^{[0,0]}$ is
the usual parity matrix used to enforce the correct commutation
relations on the DMRG matrix operators. We can understand the
unfamiliar matrix elements of the annihilation operator $c^{[-1,1/2]}$
from the Wigner-Eckart theorem: when taking the hermitian conjugate
of an operator $T$ that transforms as some representation of $SU(2)$,
what we really mean is to find the operator $T^\dagger$ such that
$\langle j^{\prime} m^{\prime} \vert T^{\dagger J}_M \vert j^{} m \rangle
= (-1)^{J-M} \langle j m \vert T^J_M \vert j^\prime m^\prime \rangle^*$.
Due to the action of the Clebsch-Gordan coefficients in the Wigner-Eckart
theorem, this is not the same as simply taking the Hermitian conjugate
of the reduced matrix operator itself. Our choice is to make the full
matrix elements of $c^{[-1,1/2]}$ identical with the usual $S^z$
representation - the matrix elements of $c^{\dagger[-1,1/2]}$ then
follow. The calculation of the matrix elements of the tensor product
of two operators acting on different sites is given by the angular
momentum theory as
\begin{equation}
\begin{array}{c}
\langle n^\prime j^\prime (\alpha^\prime_a \alpha^\prime_2
n^\prime_1 n^\prime_2 j^\prime_1 j^\prime_2)
\vert \vert \left[ \vektor{T}^{[N_1, k_1]} \otimes
\vektor{U}^{[N_2, k_2]} \right]^{[N,k]}
\vert \vert n j (\alpha_a \alpha_2 n_1 n_2 j_1 j_2) \rangle
~ \\
~ \\
= \: \left[
\begin{array}{ccc}
n_1  & n_2  & n \\
N_1  & N_2  & N \\
n\prime_1 & n^\prime_2 & n^\prime \\
\end{array}
\right]_{U(1)} \:
\left[
\begin{array}{ccc}
j_1  & j_2  & j \\
k_1  & k_2  & k \\
j\prime_1 & j^\prime_2 & j^\prime \\
\end{array}
\right]_{SU(2)}
\\
~ \\
\times \: \langle n^\prime_1 j^{\prime}_1 (\alpha^\prime_1) \vert \vert
\vektor{T}^{[N_1,k_1]} \vert
\vert n_1 j_1 (\alpha_1) \rangle \:
\langle n^\prime_2 j^\prime_2 (\alpha^\prime_2)
\vert \vert \vektor{U}^{[N_2, k_2]} \vert
\vert n_2, j_2 (\alpha_2) \rangle \; .
\end{array}
\label{eq:TensorProduct}
\end{equation}
We have written this in a slightly convoluted way so as to make
explicit the algebraic structure. The coupling coefficients for
$U(1)$ are so trivial that they are not usually represented in this
form. Indeed,
\begin{equation}
\left[
\begin{array}{ccc}
n_1  & n_2  & n \\
N_1  & N_2  & N \\
n\prime_1 & n^\prime_2 & n^\prime \\
\end{array}
\right]_{U(1)}
= \;
\delta_{n^\prime_1, N_1+n_1} \delta_{n^\prime_2, N_2 + n_2}
\delta_{n^\prime, n+N}
\delta_{n, n_1+n_2} \delta_{N, N_1+N_2}
\delta_{n^\prime, n^\prime_1+n^\prime_2}
\; ,
\end{equation}
so the $U(1)$ component of Eq.\ (\ref{eq:TensorProduct}) reduces
to exactly the coupling used in traditional DMRG. The point of
departure from traditional DMRG is that the coupling coefficients
for $SU(2)$ are not so trivial:
\begin{equation}
\left[
\begin{array}{ccc}
j_1  & j_2  & j \\
k_1  & k_2  & k \\
j\prime_1 & j^\prime_2 & j^\prime \\
\end{array}
\right]_{SU(2)}
= \;
\sqrt{2j_1^\prime+1} \sqrt{2j_2^\prime+1}  \sqrt{2j+1} \sqrt{2k+1}
\left\{
\begin{array}{ccc}
j_1  & j_2  & j \\
k_1  & k_2  & k \\
j\prime_1 & j^\prime_2 & j^\prime \\
\end{array}
\right\}
\; ,
\end{equation}
where the $\left\{ \cdots \right\}$ term is the Wigner $9j$ coefficient.

We now write the Hamiltonian
so that the operators transform as representations of
the global symmetry group. For the t-J model, this is
\begin{equation}
H = -\sqrt{2} t \sum_{\langle i,j \rangle}
(c^{\dagger [1,1/2]}_i \cdot c^{[-1,1/2]}_j + \mbox{H.c.})
- \sqrt{3} J \sum_{\langle i,j \rangle} S^{[0,1]}_i \cdot S^{[0,1]}_j
- \frac{1}{4} J \sum_{\langle i,j \rangle} n^{[0,0]}_i \cdot n^{[0,0]}_j \; .
\end{equation}
By definition, the Hamiltonian itself transforms as the $[0,0]$
identity representation. We note that
the interaction between two nearest neighbour sites $(i,j)$ in
the non-Abelian representation requires summing 4 distinct terms,
$c^\dagger_i c_j$, $c_i c^\dagger_j$, $S_i S_j$ and $n_i n_j$. In
the $S^z$ basis, there are 8 terms:
$c^\dagger_{\uparrow i} c_{\uparrow j}$,
$c_{\uparrow i} c^\dagger_{\uparrow j}$,
$c^\dagger_{\downarrow i} c_{\downarrow j}$,
$c_{\downarrow i} c^\dagger_{\downarrow j}$,
 $S^+_i S^-_j$, $S^z_i S^z_j$, $S^-_i S^+_j$ and $n_i n_j$.
Thus, although the matrix elements of the single site operators
are more difficult to calculate using the non-Abelian formulation,
there are correspondingly fewer matrix elements and operators required.

We have applied this DMRG algorithm to the two dimensional t-J
model by unrolling the two dimensional lattice into a one dimensional
model with long range interactions, following the 'worm' approach
(Liang and Pang 1994). We use periodic boundary conditions in the
y direction and open boundary conditions in the (generally longer)
x direction. Ideally, we would like to perform the calculations
with periodic boundary conditions in both directions, but this would
substantially increase the number of states required. Although the
resulting one dimensional 'worm' model is reflection symmetric at
the midpoint of the lattice, it is difficult to make use of this
symmetry in DMRG calculations due to the non-uniform
nature of the ground state. As a DMRG sweep progresses from one
end of the system towards the centre point,
the holes and spins tend to distribute themselves
in a slightly asymmetric way between the left and
right halves of the system, so that when the centre point
is reached the left block basis is biased towards
states that have too few holes, and the right block basis
is biased towards states that have too many holes
(or vice versa). Enforcing reflection symmetry by using
only one block plus its spatial reflection thus leads
to a catastrophic reduction in the number of admissible
superblock states, and a corresponding jump in the energy at
that DMRG iteration. In all two dimensional models, constructing
the initial state for the finite DMRG sweeps is more problematic
than on one dimension. In one dimension, the so-called 'infinite'
DMRG algorithm produces quite good results by itself, and usually
provides a ground state that is qualitatively very similar to the
converged finite DMRG state. In two dimensions however, the reduced
translation symmetry of the unrolled one dimensional worm means
that a genuine 'infinite' style algorithm would be rather difficult
\footnote{Although there are interesting possibilities, see for
example, Henelius (1999).}.
While there are many possible ways to construct the initial
blocks, we use the simple approach of constructing the initial
blocks 'in place'; that is, starting from an initial 4 site
system consisting of the 2 extreme sites from the left and
right ends of the worm and adding two sites at a time,
one from each end of the worm, working towards the centre
of the system. This means that for most of the warm-up sweeps,
there are no interaction terms between the left and right blocks.
An alternative procedure is to rotate the system 90 degrees, so
that the opposite ends of the worm are connected on the periodic
boundary. However this introduces many more interactions between
the left and right blocks throughout the calculation, which impacts
on the accuracy. Another possibility is to introduce extra interactions
for the warm-up sweep only, but we have not yet investigated this option.
With no interactions between the two blocks, the eigenstates of the block
density matrix coincide with the eigenstates of the block Hamiltonian.
Thus there will only be a single non-zero density matrix eigenvalue
(more, if the ground state is degenerate). The effect is that, until
the first inter-block interaction appears, $m-1$ of the block eigenstates
are essentially random vectors. The initial state can be specified
further by manipulating the target state as a function of system
size. We have done this to obtain various initial
condition: a state with all holes uniformly distributed, a phase
separated state, and several random states.

We have made calculations for various lattice sizes, keeping up
to 1200 basis states per block. Table 1 shows a comparison of the
ground state energy as a function of the number of basis
states kept, using the $U(1) \otimes U(1)$ and $U(1) \otimes SU(2)$
basis, for a typical point in the 'striped' regime (White and
Scalapino 1999; Xiang, Lou and Su 2001).
The $SU(2)$ symmetry provides a saving of a factor of
two in the number of block states required. This is very
significant as the computational complexity of the
DMRG algorithm scales as at least $O(m^3)$. However, even
with 1200 states kept in the $U(1) \otimes SU(2)$ basis
(equivalent to around 2500 states in the $U(1) \times U(1)$ basis),
the achieved energy is around $0.25\%$ higher than the estimated
true ground state energy. This compares very poorly with the
accuracies generally achieved by DMRG for one dimensional models.

In DMRG, the ground state wavefunction is iteratively improved,
but only locally. This can lead to a situation where the DMRG
converges self-consistently to an incorrect state, depending
on the initial conditions and the details of the algorithm
(Scalapino and White 2000). In many cases,
for a small number of states (but still relatively large
compared with traditional DMRG studies) we have observed
qualitatively different DMRG wavefunctions, depending on
how we perform the warm-up sweep. We have also observed
different converged wavefunctions even with the same
initial condition, simply by varying the the rate at
which the number of basis states per block is increased
as the DMRG sweeps progress. For example, with the $16 \times 6$
system used in Table 1, using 500 basis states in the
$U(1) \otimes SU(2)$ basis, the ground state is most likely
a two stripe configuration in agreement with
(White and Scalapino 1999). However, if we increase the number
of retained states at a faster rate so that it takes fewer sweeps
to reach the final total, we actually obtain a three stripe
configuration. It is not until we increased the number of states
to 800 that the three stripe configuration moves out of this local
minima and formed the two stripe configuration. Simply doing more
DMRG sweeps with 500 states is not effective.

A possible way of dealing with this problem is to compare the
energies of the competing DMRG ground states. The problem with
this approach is that DMRG only provides a variational upper
bound on the energy, and that the goodness of the variational
energy (and therefore the truncation error associated with the
DMRG state) can depend significantly on the nature of the ground
state. Thus, this requires extrapolating the energies of both
states to zero truncation error, showing that the difference
in energy is statistically significant. This is a very time consuming
procedure. In this paper, we only report results where we have a
unique ground state, independent of the initial conditions (at
least for the initial conditions we have used). However, the
problem of multiple candidate ground states depending on the
initial conditions deteriorates rather quickly as the system
size increases. Indeed, one doesn't need to increase the system
size too far before the DMRG fails to converge to a believable
ground state at all, at least in a reasonable number of sweeps.

Fig.\ 1 shows the hole density along the x direction for a fixed
number of holes, as the system size is increased. As the size grows,
the stripes tend to move further apart while keeping approximately
the same width, although it is difficult to make any real conclusions
about the stripe width from this limited data. In particular, the
width of the fluctuations in the real space density may be much
larger than the correlation width of the stripe itself, if the
stripe is delocalized. That depends on how much effect the boundary
conditions have in pinning the stripes. Fig.\ 2 shows the effect
of reducing the number of holes to 6, for the $16 \times 6$ case.
The ground state we observed is curious in that it breaks spatial
reflection symmetry, but it provides evidence that it is
energetically favourable to form a 'normal' stripe of hole
density 4/6 and one 'proto' stripe, of hole density 2/6, rather
than two stripes of equal hole density, or a single stripe. This
suggests, as with the results from Fig.\ 1, that that the
thermodynamic hole density per stripe is a constant
(which depends on J/t).

In Fig.\ 3, we attempt to find the optimal filling per stripe,
by increasing the system size and number of holes by 50\%, to a
$24 \times 6$ lattice with 12 holes. Since we consistently obtain
three stripes, this limits the hole density per unit length of
the stripes to $0.5 < d < 1$. This is the largest system that we
could study, while still being reasonably certain that the obtained
ground state is substantially independent of the initial conditions.
Thus, for these parameters and boundary conditions, the two
dimensional t-J model almost certainly has a striped ground
state, and doping the system changes the density of stripes
while the number of holes per stripe remains constant.
It is difficult, however, to extrapolate these results to make
definite conclusions about the nature of the ground state
of the t-J model in the thermodynamic limit. Because of
the half-periodic boundary conditions, the hole density is
constant in the y direction. Thus any fluctuation in the hole
density across the system, pinned by the open boundary
in that direction, will appear as a vertical stripe in
these calculations. Other possible groundstates of the thermodynamic
t-J model, such as diagonal stripes or antiferromagnetic bubbles,
are not permitted by construction. We also note that in the phase
separated region, the holes are attracted to the open boundary of
the finite system. On the other hand, in the striped regime, the
holes are repelled by the boundary. Thus the open boundary may
well have a significant effect on the nature of the ground state
that we observe, and especially, on the critical value of $J/t$
that separates stripe formation from phase separation.

It is important to emphasize that stripe formation always implies
the presence of antiphase boundaries, i.e., antiphase domain walls
in the antiferromagnet. In all approaches, antiphase boundaries
are always found in conjunction with stripes: from a simple mean-field
calculations (Zaanen and Gunnarsson 1989) to the more sophisticated
quantum numerical approaches (White and Scalapino 1998;
Morais-Smith {\sl et al.} 1998; Pryadko {\sl et al.} 1999;
Martin {\sl et al.} 2000). From our numerical data
we cannot conclude that the antiphase boundaries are a consequence
of stripes or vice-versa. What we are certain of is that the presence
of the antiphase boundaries is a clear evidence of stripe existence.
This favours our earlier observation that stripes are different
from phase separation.

An interesting explanation of the antiphase boundaries was suggested
by Nagaev (1995) \footnote{For a review, see also Markiewicz (1997)
and references cited therein.}. This brings us back to the general
theory of metal-insulator transition, as formulated by Mott (1984),
who pointed out that the number of free carriers should jump
discontinuously at the transition. Hence, there has to be a region
of phase separation near the metal insulator transition. The existence
of phase separation associated with doping away from an antiferromagnetic
phase was recognized prior to the discovery of high temperature
superconductors (Visscher 1974; Nagaev 1983).

Nagaev refers to this phase as {\sl nanoscale} phase separation
(Nagaev 1995; Markiewicz 1997), where phase separation is accompanied
by charge separation, which in a perfect isotropic crystal, form an
almost periodic structure (Nagaev 1995). Thus,
a nanoscale phase separation is realized as a form of charge
density wave. In this language the antiphase boundaries appear
as the ferromagnetic droplets (Nagaev 1983, 1995):
a hole can delocalize over a finite domain by flipping the spins
of the neighbouring Cu ion, forming a small ferromagnetic -
ferron (Nagaev 1983, 1995) islands. It is interesting to
note that if this is the origin of the antiphase boundaries then
a second hole can lower its energy by localizing on the same
ferron island, i.e., on the same antiphase boundary. This has
the appearance of a real space pairing mechanism.

Hence, looking at the stripes as being a nanoscale phase separation,
the antiphase boundaries (periodic ferrons) will appear as a consequence
of the periodic hole structure (stripes). At mean-field level this
can be understood by recalling that the superposition (coexistence) of
a charge and spin density-wave will always gives rise to ferromagnetism
(Volkov, Kopaev and Rusianov 1973; Gulacsi and Gulacsi 1986).

Work in Australia was supported by the Australian Research Council.  One
of us (I. P. McCulloch) acknowledges the hospitality of Los Alamos
National Laboratory where much of the numerical work was performed.

\newpage

\section*{References}

Anderson, P. W., 1987, Science {\bf 235}, 1196.

Barnes, S. E., 1989, Phys. Rev. B {\bf 40}, 723.

Biedenharn, L. C. and Louck, J. D., 1981, {\it Angular Momentum in
Quantum Physics}, Addison-Wesley, Massachusetts.

Blatt, J. M. and Weisskopf, V. F., 1979, {\it Theoretical
Nuclear Physics}, Springer-Verlag, New York.

Emery, V. J., Kivelson, S. A., and Lin, H. Q., 1990,
Phys. Rev. Lett. {\bf 64}, 475.

Gulacsi, M. and Gulacsi, Zs., 1986, Phys. Rev. B{\bf 33}, 6147.

Hellberg, C. S. and Manousakis, E., 1999, Phys. Rev. Lett. {\bf 83}, 132.

Henelius, P., 1999,  Phys. Rev. B{\bf 60}, 9561.

Itoyama, H., McCoy, B. M. and Perk, J. H. H., 1990,
Int. Jour. Mod. Phys. B{\bf 4}, 295.

Kivelson, S.A., Emery, V.J., and Lin, H. Q., 1990, Phys. Rev.
Lett. {\bf 64}, 475.

Liang, S. and Pang, H., 1994, Phys. Rev. B{\bf 49}, 9214.

Markiewicz, R. S., 1997, J. Phys. Chem. Solids {\bf 58}, 1179.

Martin, G. B., Gazza, C., Xavier, J. C., Feiguin, A. and Dagotto,
E., 2000, Phys. Rev. Lett. {\bf 84}, 5844.

McCulloch, I. P. and Gulacsi, M., 2000, cond-mat/0012319.

Morais-Smith, M., Dimashko, Y. M., Hasselmann, N. and Caldeiro,
A. O., 1998, Phys. Rev. B{\bf 58}, 453.

Mott, N. F., 1984, Phil. Mag. B{\bf 50}, 161.

Nagaev, E. L., 1983, {\it Physics of Magnetic Semiconductors},
Mir, Moscow.

Nagaev, E. L., 1995, Physics - Uspekhi {\bf 38}, 497.

Proceedings, 1998, {\it Conference on Spectroscopies in Novel
Superconductors}, J. Phys. Chem. {\bf 59}, No. 10 - 12.

Pryadko, L. P., Kivelson, S. A., Emery, V. J., Bazaliy, Y. B. and
Demler, E. A., 1999, Phys. Rev. Lett. {\bf 60}, 7541.

Scalapino, D. J. and White, S. R., 2000, cond-mat/0007515.

Sierra, G. and Nishino, T., 1997, Nucl. Phys. B{\bf 495}, 505.

Visscher, P. B., 1974, Phys. Rev. B{\bf 10}, 943.

Volkov, B. A., Kopaev, Y. V. and Rusianov, A. I., 1973, Zh. Eksp.
Teor. Fiz. {\bf 65}, 1984 [Sov. Phys. - JETP {\bf 38}, 991 (1974)].

White, S. R., 1992, Phys. Rev. Lett. {\bf 69}, 2863.

White, S. R., 1993, Phys. Rev. B{\bf 48}, 10, 345.

White, S. R. and Scalapino, D. J., 1999, cond-mat/9907243.

Xiang, T., Lou, J. and Su, Z., 2001, cond-mat/0102200

Yosida, K, 1980, {\it Functional Analysis}, Springer-Verlag, Berlin.

Zaanen, J. and Gunnarsson, O., 1989, Phys. Rev. B {\bf 40}, 7391.

Zaanen, J., Horbach, M. L., van Saarloos, W., 1996, Phys. Rev. B
{\bf 53}, 8671.

Zhang, F. C. and Rice, T. M., 1988, Phys. Rev. B{\bf 37}, 3759.

\table{
\caption{
Comparison of $U(1)$ and $SU(2)$ bases for the number of states versus
ground state energy of a $16 \times 6$ t-J system with $J = 0.35$,
$t = 1$, 8 holes and cylindrical boundary conditions. The results using
the $U(1)$ basis are from reference (White and Scalapino 1999)
We also include an estimate of the true energy, extrapolated to zero
truncation error.
}
\[
\begin{array}{rr r@{.}l}
\hline \hline
\mbox{basis} & \multicolumn{1}{c}{m} & \multicolumn{1}{c}{E} \\ \hline
U(1)         & 1000 & -52 & 279        \\
SU(2)        & 500 & -52 & 284 \\
SU(2)        & 800 & -52 & 463 \\
SU(2)        & 1200 & -52 & 520 \\
\multicolumn{1}{c}{-} & \infty & -52 & 65 \pm 0.05 \\
\hline \hline
\end{array}
\]
}

\newpage

\section*{Figure Captions}

Figure 1. Hole density in the $x$ direction for lattice sizes $16 \times 6$, $18 \times 6$
and $20 \times 6$, 8 holes, at $J/t = 0.35$.

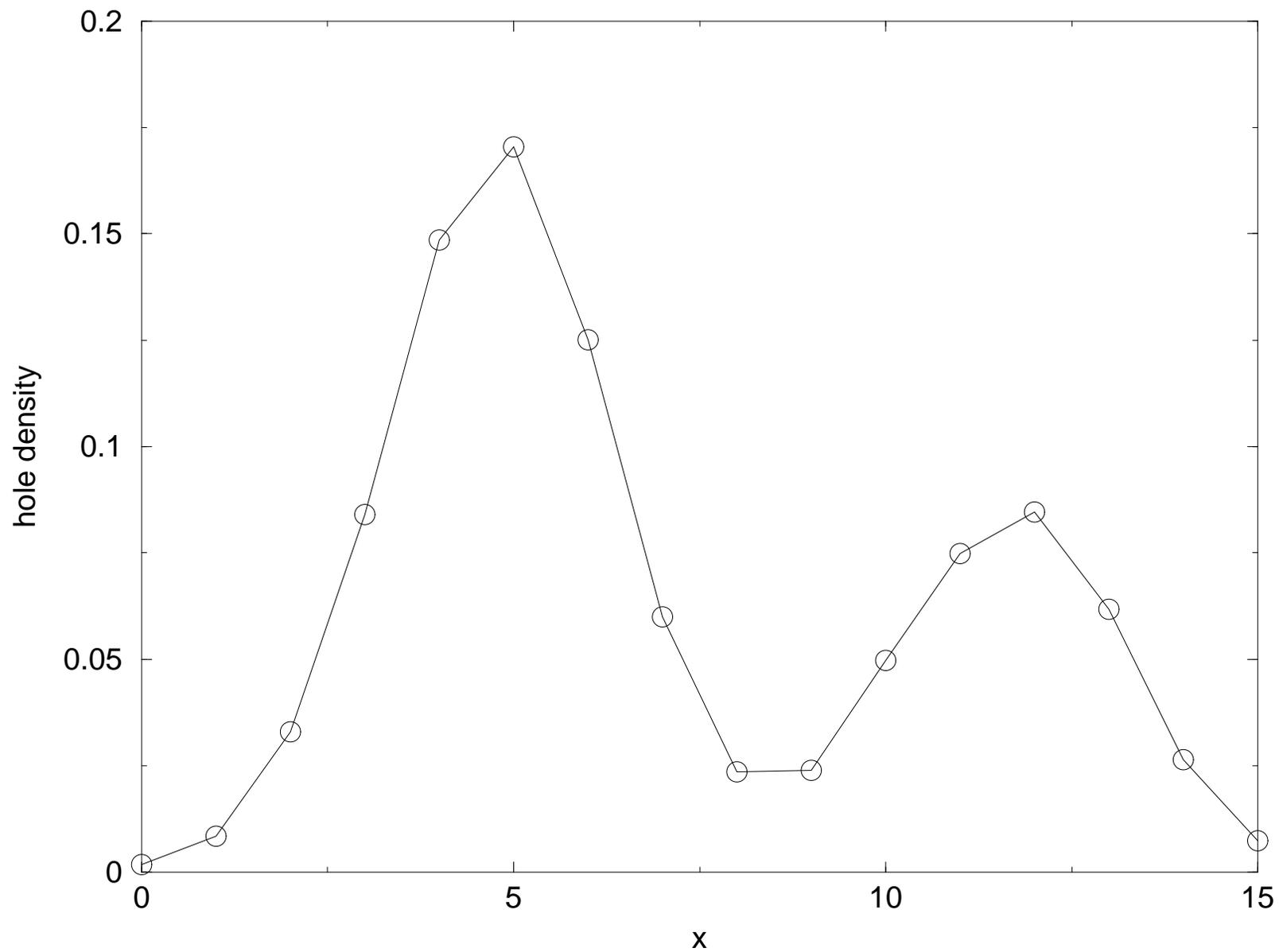
Figure 2. Hole density in the $x$ direction for lattice size $16 \times 6$, 6 holes,
$J/t = 0.35$.

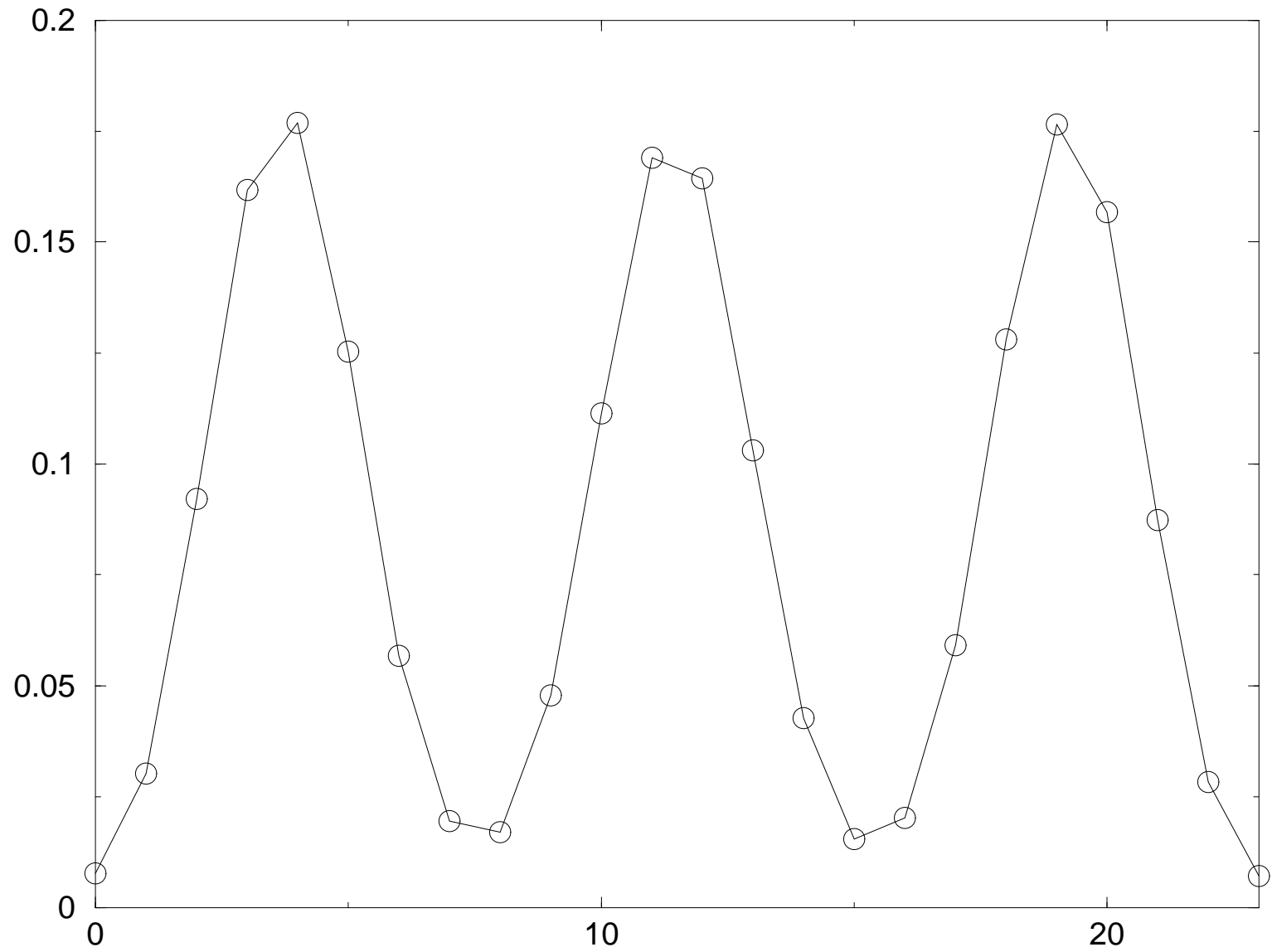
Figure 3. Hole density in the $x$ direction for lattice size $24 \times 6$, 12 holes,
$J/t = 0.35$.

\end{document}